\documentclass[aps,prl,10pt,superscriptaddress,amsmath,twocolumn,floatfix]{revtex4}
\usepackage{pstricks}
\usepackage{amsmath}
\usepackage{amsfonts}
\usepackage{graphicx}
\usepackage{color}

\newcommand{\parti}[2]{\frac{\partial #1}{\partial #2}}

\newcommand{\avg}[1]{\langle#1\rangle}

\newcommand{\bs}[1]{\boldsymbol {#1}}
\begin{document}

\title{Evading quantum mechanics}

\author{Mankei Tsang}
\email{eletmk@nus.edu.sg}
\affiliation{Department of Electrical and Computer Engineering,
  National University of Singapore, 4 Engineering Drive 3, Singapore
  117583}

\affiliation{Department of Physics, National University of Singapore,
  2 Science Drive 3, Singapore 117551}

\affiliation{Center for Quantum Information
and Control, University of New Mexico, MSC07-4220, Albuquerque, New
Mexico 87131-0001, USA}

\author{Carlton M.~Caves}
\affiliation{Center for Quantum Information and Control, University
of New Mexico, MSC07-4220, Albuquerque, New Mexico 87131-0001, USA}
\affiliation{Centre for Engineered Quantum Systems, School of Mathematics
and Physics, The University of Queensland, St.~Lucia, Brisbane 4072, Australia}

\date{\today}

\begin{abstract}
  Quantum mechanics is potentially advantageous for certain
  information-processing tasks, but its probabilistic nature and
  requirement of measurement back action often limit the precision of
  conventional classical information-processing devices, such as
  sensors and atomic clocks.  Here we show that by engineering the
  dynamics of coupled quantum systems, it is possible to construct a
  subsystem that evades the measurement back action of quantum
  mechanics, at all times of interest, and obeys any classical
  dynamics, linear or nonlinear, that we choose.  We call such a
  system a \textit{quantum-mechanics-free subsystem\/} (QMFS).  All of
  the observables of a QMFS are quantum-nondemolition (QND)
  observables; moreover, they are dynamical QND observables, thus
  demolishing the widely held belief that QND observables are
  constants of motion.  QMFSs point to a new strategy for designing
  classical information-processing devices in regimes where quantum
  noise is detrimental, unifying previous approaches that employ QND
  observables, back-action evasion, and quantum noise cancellation.
  Potential applications include gravitational-wave detection,
  optomechanical force sensing, atomic magnetometry, and classical
  computing.  Demonstrations of dynamical QMFSs include the generation
  of broad-band squeezed light for use in interferometric
  gravitational-wave detection, experiments using entangled atomic
  spin ensembles, and implementations of the quantum Toffoli~gate.
\end{abstract}

\maketitle

Scientists studying gravitational-wave detection, concerned that
gravitational-wave detectors would be limited by quantum
uncertainties, realized that if a quantum observable, represented by a
self-adjoint operator $O(t)$ in the Heisenberg picture, commutes with
itself at times $t$ and $t'$ when the observable is measured, viz.,
\begin{align}
[O(t),O(t')] &= 0\;,
\label{qnd}
\end{align}
then measurement back action does not limit accurate measurement of
this observable or any classical signal coupled to it.  An observable
that obeys Eq.~(\ref{qnd}) is called a \textit{quantum-nondemolition
  (QND) observable\/}~\cite{caves,unruh,bvt}.  An observable that
satisfies Eq.~(\ref{qnd}) at all times is called a
\emph{continous-time\/} QND observable, and one that satisfies
Eq.~(\ref{qnd}) only at discrete times is called a
\emph{stroboscopic\/} QND observable.

The most well-known QND observables are ones that remain
\emph{static\/} in the absence of classical signals, viz.,
\begin{align}
O(t) &= O(t')\;.
\label{constant}
\end{align}
Peres showed that Eq.~(\ref{constant}) is indeed a necessary condition
for an observable to be QND in continuous time \emph{if} $O(t)$ has a
discrete spectrum (and has no explicit time dependence in the
Schr\"odinger picture)~\cite{peres_prd}.  Nowadays it is often assumed
that Eqs.~(\ref{qnd}) and~(\ref{constant}) are interchangeable as the
QND condition~\cite{peres,wiseman,monroe}.  Overemphasis on
Eq.~(\ref{constant}) as the QND condition trivializes the QND concept
and has even led to calls for its retirement~\cite{monroe}.

An assumption that Eq.~(\ref{constant}) is a necessary QND condition
implies that measurement back action would always introduce additional
uncertainties to any quantum system with richer dynamics than
Eq.~(\ref{constant}) and limit one's ability to process classical
information accurately.  A famous example of such thinking is the
standard quantum limit to force sensing~\cite{braginsky,caves}, which
arises from back-action noise and was considered to be a fundamental
limit on force sensitivity with position measurements.

The central result of this paper is to show that there exists a much
wider class of observables that obey the QND condition~(\ref{qnd}).
To this end, we generalize the concept of a QND observable to that of
a \emph{quantum-mechanics-free subsystem} (QMFS)~\cite{dfs}, which is
a set of observables $\mathcal{O}=\{O_1,O_2,\dots,O_{N}\}$ that obey,
in the Heisenberg picture,
\begin{align}
[O_j(t),O_k(t')] = 0
\textrm{\ for all $j$ and $k$,}
\label{cfs}
\end{align}
at all times $t$ and $t'$ when the observables are
measured. Mathematically, Eq.~(\ref{cfs}) guarantees the classicality
of a QMFS by virtue of the spectral theorem, which allows one to map
the commuting Heisenberg-picture operators to processes in a classical
probability space~\cite{reed_simon,belavkin,bouten}.

The correspondence between QND observables and classical processes
implies that a QMFS is immune to the laws of quantum mechanics, such
as the Heisenberg uncertainty principle and measurement invasiveness,
\emph{at all times of interest}.  This exact classicality of a QMFS
should be contrasted with the approximate classicality that emerges in
the macroscopic limit through coarse graining or
decoherence~\cite{zurek,decoh,milburn}.

Because any subset of a QMFS is also a QMFS, a decohering quantum
system can contain a QMFS as well, if a set of system operators
together with the environment operators form a larger \hbox{QMFS}.
The environment then behaves as classical dissipation and fluctuation
in the accessible part of the \hbox{QMFS}. In the same vein, we can
broadly define measurements of any strength as QND if they can be
dilated to projective measurements on a larger QMFS.  Repeated QND
measurements, even if they are projective on the system of interest,
need not reproduce the same outcomes, as the QMFS can evolve during
the measurements.

\begin{figure}[htbp]
\includegraphics[width=0.45\textwidth]{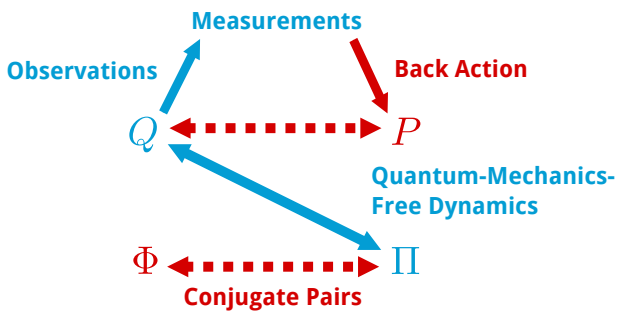}
\caption{(Color). A quantum-mechanics-free subsystem (QMFS, in blue),
  which consists of dynamically coupled quantum-nondemolition (QND)
  observables $\{Q,\Pi\}$.  The QMFS naturally evades measurement back
  action; measurements of $Q$, for example, produce back action onto
  the conjugate observable $P$, which does not influence the QMFS
  observables $Q$ and $\Pi$.}
\label{fig:qmfs1}
\end{figure}

A continuous-time dynamical QMFS must consist of continuous variables,
given Peres's result~\cite{peres_prd}.  To construct such a system,
consider two sets of canonical positions and momenta,
$\{Q,P\}=\{Q_1,Q_2,\dots,Q_M,P_1,P_2,\dots,P_M\}$ and
$\{\Phi,\Pi\}=\{\Phi_1,\Phi_2,\dots,\Phi_M,\Pi_1,\Pi_2,\dots,\Pi_M\}$,
which obey the canonical commutation relations,
\begin{align}
[Q_j,P_k] &= [\Phi_j,\Pi_k] = i\hbar \delta_{jk}\;,
\end{align}
and otherwise commute with one another.  Suppose the Hamiltonian has the form
\begin{align}
H &= \frac{1}{2}\sum_{j=1}^M (P_j f_j+f_jP_j + \Phi_j g_j+g_j\Phi_j) + h\;,
\label{hamiltonian}
\end{align}
where $f_j=f_j(Q,\Pi,t)$, $g_j=g_j(Q,\Pi,t)$, and $h=h(Q,\Pi,t)$ are
arbitrary, Hermitian-valued functions.  The equations of motion for
$Q_j(t)$ and $\Pi_j(t)$ in the Heisenberg picture become
\begin{align}
\dot Q_j &= f_j\bigl(Q(t),\Pi(t),t\bigr)\;,
&
\dot\Pi_j &= -g_j\bigl(Q(t),\Pi(t),t\bigr)\;.
\label{Pmotion}
\end{align}
The $Q$ and $\Pi$ variables are dynamically coupled to each other, but
not to the incompatible set $\{\Phi,P\}$, and thus obey
Eq.~(\ref{cfs}) and form a \hbox{QMFS}, as depicted by
Fig.~\ref{fig:qmfs1}.

These QMFS variables can follow arbitrary classical trajectories in
continuous time, including ones that are perturbed by classical
signals or do not obey classical Hamiltonian dynamics.  The QMFS
variables can be prepared with arbitrarily small quantum
uncertainties, or when monitored with sufficient accuracy, they will
tend to such small quantum uncertainties.  The measurement back action
acts on the conjugate variables $\{\Phi,P\}$.  The resulting large
quantum uncertainties required by the Heisenberg uncertainty principle
are isolated in the variables $\{\Phi,P\}$, which do not influence the
\hbox{QMFS}.

The classical trajectories followed within the QMFS do not have to
obey Hamiltonian dynamics, but they will be those of a classical
Hamiltonian $\tilde H(Q,\Pi,t)$ if we choose
\begin{align}
f_j&=\parti{\tilde H}{\Pi_j}=\dot Q_j\;,
&
g_j&=\parti{\tilde H}{Q_j}=-\dot\Pi_j\;.
\end{align}
This QMFS was first suggested by Koopman as a formulation of classical
Hamiltonian dynamics in a Hilbert space~\cite{koopman,peres}, but its
application to back-action evasion for quantum systems has not
hitherto been appreciated.

A prime example of this sort of QMFS arises in the case of two pairs
of canonical variables ($M=1$) when the QMFS dynamics is that of a
harmonic oscillator with mass $m$ and frequency $\omega$, i.e.,
classical Hamiltonian $\tilde H=\Pi^2/2m+m\omega^2Q^2/2$ and QMFS
equations of motion
\begin{align}\label{qmfs1}
\dot Q&=\frac{\Pi(t)}{m}\;,
&
\dot\Pi&=-m\omega^2 Q(t)\;.
\end{align}
The overall quantum dynamics is that of the quadratic Hamiltonian
\begin{align}\label{qmfs2}
H &= \frac{P\Pi}{m} + m\omega^2 \Phi Q\;,
\end{align}
so $\{\Phi,P\}$ form an identical harmonic-oscillator QMFS in this case.

\begin{figure}[htbp]
\includegraphics[width=0.4\textwidth]{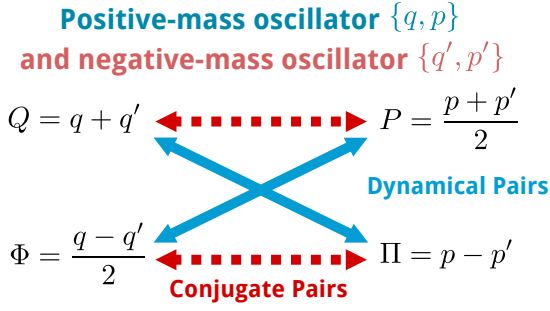}
\caption{(Color). Given two pairs of canonical variables, $\{q,p\}$
  and $\{q',p'\}$, the collective position $Q$ and average momentum
  $P$ form a conjugate pair, and the relative position $\Phi$ and
  relative momentum $\Pi$ form another conjugate pair.  The conjugate
  pairs are restricted by the Heisenberg uncertainty principle.  If
  the unprimed and primed variables are those of a positive-mass and
  negative-mass oscillator, however, $\{Q,\Pi\}$ and $\{\Phi,P\}$ form
  dynamical oscillator pairs, each of which is a QMFS.  Reducing
  uncertainty in one QMFS at the expense of the other is equivalent to
  introducing EPR correlations among the original oscillator
  variables.}
\label{fig:qmfsa}
\end{figure}

To get an idea of what this overall Hamiltonian means and how it might
be implemented, we transform to new canonical variables
\begin{align}\label{qmfs3}
Q &= q+q'\;,
&P &= \frac{p + p'}{2}\;,
\\
\Phi &= \frac{q-q'}{2}\;,
&\Pi &= p-p'\;,
\end{align}
in terms of which the Hamiltonian~(\ref{qmfs2}) becomes
\begin{align}\label{qmfs4}
H &=\frac{p^2}{2m} + \frac{1}{2}m\omega^2 q^2
- \frac{p'^2}{2m}-\frac{1}{2}m\omega^2 q'^2\;.
\end{align}
The transformed quantum system consists of two harmonic oscillators, one with positive mass and the other (primed) with negative mass.  The relationships among the two sets of variables are summarized in Fig.~\ref{fig:qmfsa}.

A negative-mass oscillator is not the same as a particle moving in an
inverted potential.  Instead, the entire Hamiltonian is inverted.  The
dynamics consists of oscillations at frequency $\omega$, just as for a
positive-mass oscillator, and the energy eigenstates are the same as
those of a positive-mass oscillator, but the ladder of energy levels
runs down instead of up; each quantum of excitation reduces the
oscillator energy by $\hbar\omega$.

\begin{figure}[htbp]
\includegraphics[width=0.48\textwidth]{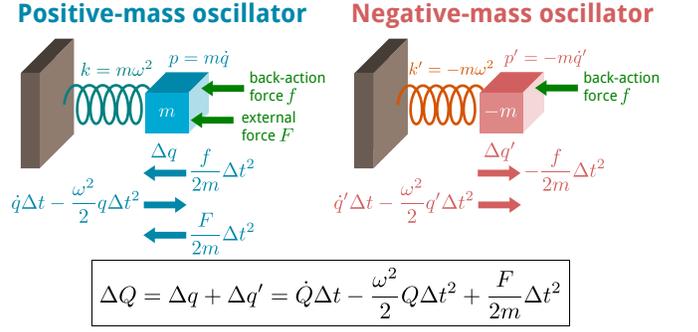}
\caption{(Color). Behavior of positive- and negative-mass oscillators
  during a short time interval $\Delta t$.  Monitoring the collective
  position $Q$ leads to the same back-action force $f$ on both
  oscillators.  The positive-mass oscillator is ``pushed'' by $f$,
  whereas the negative-mass oscillator is ``pulled'' in the opposite
  direction; the effect of the back-action thus cancels in $Q$.  The
  evolution of $Q$ under an external force $F$ applied to the
  positive-mass oscillator is exactly the same as a single oscillator.
  Positive- and negative-mass oscillators can be realized as blue and
  red sidebands of a carrier frequency.}
\label{fig:qmfsb}
\end{figure}

The continuous-time QMFS of Eqs.~(\ref{qmfs1})--(\ref{qmfs4}) is
naturally back-action-evading~\cite{braginsky,caves,bvt}, as
measurements of $Q(t)$ or $\Pi(t)$ introduce back action to the
conjugate variables, $P(t)$ or $\Phi(t)$, which are never coupled to
the measured subsystem.  A complementary perspective is to consider
the QMFS as a quantum-noise-cancellation scheme~\cite{qnc}:
measurements of $Q=q+q'$ produce equal back action onto $p$ and $p'$,
which cancels coherently in the dynamical variable, $\Pi=p-p'$, that
is coupled to~$Q$.  Quantum noise cancellation is illustrated in
Fig.~\ref{fig:qmfsb}.

A pairing of positive- and negative-mass oscillators occurs naturally
as mirror sidebands of a carrier frequency $\Omega$.  Thus consider
two field modes, with frequencies $\Omega\pm\omega$ placed
symmetrically about $\Omega$.  The Schr\"odinger-picture Hamiltonian
of the two modes is $H_{\rm SP}=\hbar(\Omega+\omega)a^\dagger
a+\hbar(\Omega-\omega)b^\dagger b$, where $a$ and $b$ are annihilation
operators for the blue and red sidebands.
To see this formally, we transform to the modulation
picture~\cite{caves85}, which moves the rapid oscillation at the
carrier frequency from quantum states to operators.  If we explicitly
remove this rapid oscillation from the annihilation operators,
\begin{align}
ae^{i\Omega t}&=\sqrt{\frac{\omega}{2\hbar}}\biggl(q+{\frac{ip}{\omega}}\biggr)\;,
&
be^{i\Omega t}&=\sqrt{\frac{\omega}{2\hbar}}\biggl(q'+{\frac{ip'}{\omega}}\biggr)\;,
\end{align}
we end up with a pair of oscillators that oscillate at the modulation
frequency~$\omega$.  The blue sideband is a positive-mass oscillator,
and the red sideband is a negative-mass oscillator.  The
modulation-picture Hamiltonian, $H_{\rm MP}=\hbar\omega(a^\dagger
a-b^\dagger b)$, is the two-oscillator Hamiltonian~(\ref{qmfs4}) with
$m=1$.

To illustrate the connection to the QMFS variables, it is instructive
to introduce an electromagnetic field operator given by
\begin{equation}
E=E^{(+)}+E^{(-)}=E_1\cos\Omega t+E_2\sin\Omega t\;.
\end{equation}
Here
\begin{equation}\label{Eplus}
E^{(+)}=\sqrt{\frac{\hbar\Omega}{2}}(a + b)={\frac{1}{2}}(E_1+iE_2)e^{-i\Omega t}
\end{equation}
and $E^{(-)}=E^{(+)\dagger}$ are the positive- and negative-frequency
parts of the field, and $E_1$ and $E_2$ are the field's (Hermitian)
\emph{quadrature components}, defined relative to the carrier
frequency.  The quadrature components take the form
$E_1=\sqrt{\hbar\Omega}(\alpha_1+\alpha_1^\dagger)$ and
$E_2=\sqrt{\hbar\Omega}(\alpha_2+\alpha_2^\dagger)$, where
\begin{align}
\alpha_1&={\frac{1}{\sqrt2}}(ae^{i\Omega t}+b^\dagger e^{-i\Omega t})
=\frac{1}{2}
\sqrt{\frac{\omega}{\hbar}}\biggl(Q+i{\frac{\Pi}{\omega}}\biggr)\;,
\\
\alpha_2&=-{\frac{i}{\sqrt2}}(ae^{i\Omega t}-b^\dagger e^{-i\Omega t})
=\sqrt{\frac{\omega}{\hbar}}\biggl(-i\Phi+{\frac{P}{\omega}}\biggr)
\end{align}
are the \emph{quadrature
  amplitudes\/}~\cite{caves85,schumaker85,shapiro}.  Just as the
annihilation operators $a$ and $b$ are quantum operators for the
classical variables that encode the amplitude and phase of the modal
oscillations at frequencies $\Omega\pm\omega$, so the quadrature
amplitudes $\alpha_1$ and $\alpha_2$ encode the amplitude and phase of
the oscillations of the quadrature components.  Each quadrature
amplitude describes oscillations within a~\hbox{QMFS}.

Two-mode squeezed states~\cite{caves85,schumaker85} take advantage of
this QMFS structure in an electromagnetic wave to decrease the quantum
uncertainties associated with one quadrature component, while
increasing the uncertainty associated with the other.  Homodyne
detection at the carrier frequency~\cite{shapiro} measures one
quadrature component and, hence, the signal within a~\hbox{QMFS}.  The
broadband squeezed states now being introduced into interferometric
gravitational-wave detectors~\cite{mehmet,LIGO} use two-mode squeezing
over a wide bandwidth of modulation frequencies and are thus an
example of using dynamical QMFSs in probing the motion of a mechanical
system.  In such broadband squeezed states, the $\{Q,\Pi\}$ variables
are not correlated with the $\{\Phi,P\}$ set, but the oscillator
variables for the blue and red sidebands are necessarily correlated in
the Einstein-Podolsky-Rosen (EPR) sense~\cite{braunstein}.

These considerations suggest a way to implement a QMFS using a
mechanical oscillator.  If the oscillator is probed by an optical beam
with carrier frequency $\Omega$, the negative-mass harmonic oscillator
can be simulated by an optical mode in a cavity with a red-detuned
resonance at $\Omega-\omega$~\cite{qnc}.  This strategy of introducing
an auxiliary mode to form a QMFS and measuring the collective
position $Q$ enables one to beat the standard quantum limit for force
detection~\cite{qnc,qcrb} and entangle the mechanical oscillator with the
auxiliary mode~\cite{hammerer}.

Another way of implementing Eq.~(\ref{qmfs4}) is to use two spin
ensembles, both of which have total angular momentum $J_0$.  Suppose
the ensembles are polarized nearly maximally, but oppositely along the
direction of an applied magnetic field $B_0{\bf e}_z$.  The average
angular momenta are then $\avg{\bs J}=-\avg{\bs J'}\simeq J_0{\bf
  e}_z$.  Off-axis polarizations precess about the magnetic field.
For large angular momentum, the precessional oscillations of the $x$
and $y$ components of the angular momenta are identical to the
phase-space trajectory of a harmonic oscillator.  Moreover, the
aligned angular momentum $\bs J$ has magnetic sublevels whose energy
increases away from maximal polarization, making it a positive-mass
oscillator, whereas for the anti-aligned angular momentum $\bs J'$,
the magnetic sublevels decrease in energy, making it a negative-mass
oscillator.  The resulting QMFS structure has been used to achieve
quantum noise cancellation~\cite{julsgaard,wasilewski}.

Formally, we have, in the Holstein-Primakoff approximation,
\begin{align}
[J_{x},J_{y}] &= i\hbar J_{z}\simeq i\hbar J_0\;,
\\
[J_{x}',J_{y}'] &= i\hbar J_{z}'\simeq -i\hbar J_0\;.
\end{align}
Defining canonical position and momentum operators by
\begin{align}
q=J_x/\sqrt{J_0}\;,\qquad&p=J_y/\sqrt{J_0}\;,
\\
q'=J_x'/\sqrt{J_0}\;,\qquad&p'=-J_y'/\sqrt{J_0}\;,
\end{align}
and using $J_z\simeq\sqrt{J_0(J_0+1)}-(q^2+p^2)/2$ and $J'_z\simeq-\sqrt{J_0(J_0+1)}+(q'^2+p'^2)/2$, the Hamiltonian becomes
\begin{equation}
H=-\gamma B_0(J_z + J_z')\simeq\frac{\gamma B_0}{2}(q^2 + p^2-q'^2-p'^2)\;,
\end{equation}
which has the form of Eq.~(\ref{qmfs4}).

Since $Q$ and $\Pi$ commute at all times, continuous measurements of
one reveal information about the other with no back action, and the
pair can have uncertainties that violate the Heisenberg uncertainty
principle.  As noted above, this violation means that the two physical
oscillators, $\{q,p\}$ and $\{q',p'\}$, are entangled in the EPR
sense.  The collective-angular-momentum experimental demonstration of entanglement
in~\cite{julsgaard} can thus be regarded as a demonstration of a QMFS
that behaves as a classical harmonic oscillator.  Moreover, the
magnetometer reported in~\cite{wasilewski} demonstrates the use of a
dynamical QMFS for sensing that does not suffer from
quantum-measurement back action.  The dynamical QMFS ($\{\Phi,P\}$ in
this case) has the advantage of being resonant with oscillating
magnetic-field signals in the $x$ or $y$ direction near the tunable
Larmor frequency $\gamma B_0$, whereas a static QMFS with operators
that obey Eq.~(\ref{constant}) is much less sensitive to oscillating
signals when the signal phase is unknown.

It is possible to construct discrete-variable QMFSs as well, as long
as the QND condition is imposed stroboscopically.  Examples come from
quantum computation.  Suppose we have a collection of qubits.  The
simultaneous eigenstates of the Pauli $Z$ operators for all the qubits
are specified by bit strings of eigenvalues of the $Z$ operators and
are often called the computational basis.  A quantum gate that
permutes computational-basis states executes a classical (reversible)
gate on the input bit string.  In the Heisenberg picture, such a gate
takes the input $Z$ operators to output $Z$ operators that are
functions of the inputs and thus commute with the inputs.  The
classical information processing performed by the gate can be regarded
as noiseless information processing performed within the QMFS of the
Pauli $Z$ operators restricted to times pre- and post-gate.

An example of such a gate is the controlled-NOT gate~\cite{nielsen},
which transforms the Pauli $Z$ operators according to $Z_{1}'=Z_1$ and
$Z'_2=Z_1Z_2$, where unprimed and primed operators refer to pre- and
post-gate times.  A more ambitious example is the three-qubit Toffoli
gate~\cite{nielsen,feynman}, a controlled-controlled-NOT gate, which
transforms the Pauli $Z$ operators according to
\begin{align}\label{Toffoli}
Z'_1&=Z_1\;,\qquad
Z'_2=Z_2\;,
\\
Z'_3&=\Bigl(I-\textstyle\frac{1}{2}(I-Z_1)(I-Z_2)\Bigr)Z_3\;,
\end{align}
where $I$ is the identity operator.  For both these gates, since the
output $Z$ operators commute with the input, the $Z$ operators can be
mapped to classical bits that undergo classical information processing
between input and output.  Classical Toffoli gates form a set of
universal gates for (reversible) classical computation~\cite{toffoli},
so one can construct any classical discrete-variable dynamics in
discrete time using a circuit of quantum Toffoli gates.  Thus Benioff
and Feynman's quantum-mechanical computer for universal classical
computation~\cite{benioff1,benioff2,feynman} is an example of information
processing within a dynamical \hbox{QMFS}.  Experimental
demonstrations of the quantum Toffoli gate have been reported
in~\cite{monz,fedorov,reed}.

The existence of QMFSs does not contradict proven quantum limits to
classical information processing, such as the quantum Cram\'er-Rao
bound~\cite{helstrom,holevo,qcrb}, as all such limits are derived from
quantum mechanics.  This implies that proven quantum limits should
either involve incompatible observables outside a QMFS or have
effectively classical origins.  The concept of a QMFS thus unifies
under a single framework the several strategies for evading
measurement back action, such as QND observables, back-action evasion,
and quantum noise cancellation.  Given what we have seen from the
example of force sensing, where a QMFS can beat the standard quantum
limit and saturate the quantum Cram\'er-Rao bound~\cite{qnc,qcrb}, we
envision QMFSs to be a useful tool for overcoming heuristic quantum
limits and approaching proven limits for classical information
processing applications in general.

We acknowledge useful discussions with Joshua Combes. This material is
based on work supported in part by the Singapore National Research
Foundation under NRF Grant No.~NRF-NRFF2011-07, US National Science
Foundation Grant Nos.~PHY-0903953 and PHY-1005540, and US Office of
Naval Research Grant No.~N00014-11-1-0082.

\end{document}